\documentclass[11pt,a4paper]{article}
\usepackage[hyperref]{emnlp2018}
\usepackage{times}
\usepackage{latexsym}
\usepackage{mathrsfs}
\usepackage[linesnumbered,ruled,vlined]{algorithm2e}
\usepackage[noend]{algpseudocode}
\usepackage{amsmath}
\usepackage{amssymb}
\usepackage{graphicx}
\usepackage{booktabs}
\usepackage{lscape}

\usepackage{url}

\usepackage{amsmath}
\usepackage{bm}

\aclfinalcopy 
\def\aclpaperid{2100} 

\usepackage{xcolor}

\usepackage{xspace}

\newcommand{\cparagraph}[1]{\vspace{1.5mm}\noindent\textbf{#1.}}

\DeclareMathOperator{\sigmoid}{sigmoid}

\newcommand{\entropy}{\texttt{EntropyModel}\xspace}
\newcommand{\RLunif}{\texttt{RL\_uniform}\xspace}
\newcommand{\RLpop}{\texttt{RL\_popularity}\xspace}
\newcommand{\directR}{\texttt{DirectReward}\xspace}
\newcommand{\RN}{\texttt{RewardNet}\xspace}
\newcommand{\ORN}{\texttt{ObjectRewardNet}\xspace}
\newcommand{\unifSimu}{\texttt{UnifSimulator}\xspace}
\newcommand{\popSimu}{\texttt{PopSimulator}\xspace}

\title{Playing 20 Question Game with Policy-Based Reinforcement Learning}

\author{Huang Hu$^{1}$\thanks{\ \ The work was done when the first author was an intern in Microsoft XiaoIce team.}, Xianchao Wu$^3$, Bingfeng Luo$^1$, Chongyang Tao$^{1}$,
	\\\textbf{Can Xu$^2$, Wei Wu$^2$ \and Zhan Chen$^3$}\\
	$^1$Peking University, Beijing, China\\
	$^2$Microsoft Corporation, Beijing, China \\
    $^3$Microsoft Development Co., Ltd, Tokyo, Japan \\
	{\tt $^1$\{tonyhu,bf\_luo,chongyangtao\}@pku.edu.cn} \\
	{\tt $^{2,3}$\{xiancwu,can.xu,wuwei,zhanc\}@microsoft.com} \\}
\date{}

\begin{document}
\maketitle
\begin{abstract}

The 20 Questions (Q20) game is a well known game which encourages deductive reasoning and creativity.
In the game, the answerer first thinks of an object such as a famous person or a kind of animal.
Then the questioner tries to guess the object by asking 20 questions.
In a Q20 game system, the user is considered as the answerer while the system itself acts as the questioner which requires a good strategy of question selection to figure out the correct object and win the game.
However, the optimal policy of question selection is hard to be derived due to the complexity and volatility of the game environment.
In this paper, we propose a novel policy-based Reinforcement Learning (RL) method, which enables the questioner agent to learn the optimal policy of  question selection through continuous interactions with users.
To facilitate training, we also propose to use a reward network to estimate the more informative reward.
Compared to previous methods, our RL method is robust to noisy answers and does not rely on the Knowledge Base of objects.
Experimental results show that our RL method clearly outperforms an entropy-based engineering system and has competitive performance in a noisy-free simulation environment.
\footnote{Our code and dataset is available at \url{https://github.com/stonyhu/Q20-DeepRL}}
\end{abstract}
\section{Introduction}

The 20 Question Game (Q20 Game) is a classic game that requires deductive reasoning and creativity.
At the beginning of the game, the \textbf{\emph{answerer}} thinks of a target object and keeps it concealed.
Then the \textbf{\emph{questioner}} tries to figure out the target object by asking questions about it, and the answerer answers each question with a simple ``Yes'', ``No'' or ``Unknown'', honestly.
The questioner wins the game if the target object is found within 20 questions.
In a Q20 game system, the user is considered as the answerer while the system itself acts as the questioner which requires a good question selection strategy to win the game.

As a game with the hype \emph{read your mind}, Q20 has been played since the 19th century, and was brought to screen in the 1950s by the TV show \emph{Twenty Questions}.
Burgener's program~\cite{Burgener2006Artificial} further popularized Q20 as an electronic game in 1988, and modern virtual assistants like Microsoft XiaoIce and Amazon Alexa also incorporate this game into their system to demonstrate their intelligence.

However, it is not easy to design the algorithm to construct a Q20 game system.
Although the decision tree based method seems like a natural fit to the Q20 game, it typically require a well defined Knowledge Base (KB) that contains enough information about each object, which is usually not available in practice.
Burgener~\shortcite{Burgener2006Artificial} instead uses a object-question relevance table as the pivot for question and object selection, which does not depend on an existing KB. 
Wu et al.~\shortcite{wu2018q20} further improve the relevance table with a lot of engineering tricks.
Since these table-based methods greedily select questions and the model parameters are only updated by rules, their models are very sensitive to noisy answers from users, which is common in the real-world Q20 games.
Zhao and Maxine~\shortcite{zhao2016towards} utilizes a value-based Reinforcement Learning (RL) model to improve the generalization ability but still relies on the existing KB.

In this paper, we formulate the process of question selction in the game as a Markov Decision Process (MDP),
and further propose a novel policy-based RL framework to learn the optimal policy of question selection in the Q20 game.
Our questioner agent maintains a probability distribution over all objects to model the confidence of the target object, and updates the confidence based on answers from the user.
At each time-step. the agent uses a policy network $\pi_\theta(a|\bm{s})$ to take in the confidence vector and output a question distribution for selecting the next question.
To solve the problem that there is no immediate reward for each selected question, we also propose to employ a RewardNet to estimate the appropriate immediate reward at each time-step, which is further used to calculate the long-term return to train our RL model.
Our RL framework makes the agent robust to noisy answers since the model parameters are fully learnable and the question distribution from $\pi_\theta(a|\bm{s})$ provides us with a principled way to sample questions, which enables the agent to jump out of the local optimum caused by incorrect answers and also introduces more randomness during training to improve the model generalization ability.
Furthermore, the ability to sample questions, compared to greedy selection, also improves the diversity of the questions asked by our agent, which is crucial for user experience.

Our contributions can be summarized as follows:
(1) We propose a novel RL framework to learn the optimal policy of question selection in the Q20 game without any dependencies on the existing KBs of target objects. Our trained agent is robust to noisy answers and has a good diversity in its selected questions.
(2) To make the reward more meaningful, we also propose a novel neural network on reward function approximation to deliver the appropriate immediate rewards at each time-step.
(3) Extensive experiments show that our RL method clearly outperforms a highly engineered baseline in the real-world Q20 games where noisy answers are common. Besides, our RL method is also competitive to that baseline on a noise-free simulation environment.

\section{Method}

\begin{figure}[!t]
	\centering
	\includegraphics[width=0.48\textwidth]{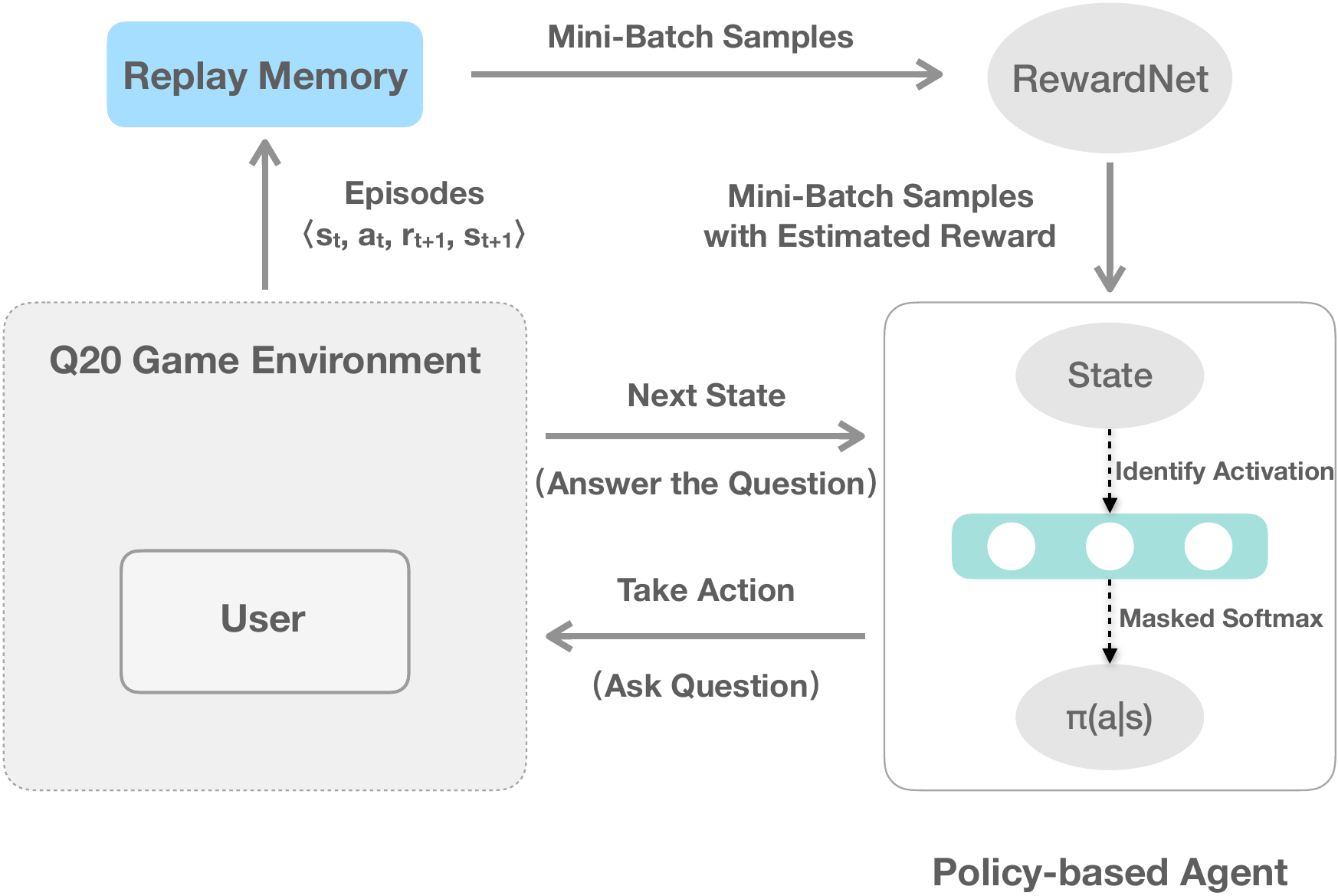}
	\vspace{-2em}
	\caption{The overview of our RL framework.}
	\vspace{-1em}
	\label{fig:method_overview}
\end{figure}

In this section, we first describe our RL framework for playing the Q20 game, which is shown in the Fig.~\ref{fig:method_overview}.
The user in our system is the answerer who thinks of a target object $o_{tgt}$ in the object set $\mathcal{O}$ at the beginning of the game.
Our policy-based agent acts as the questioner that can ask 20 questions to figure out what exactly $o_{tgt}$ is.
Specifically, an internal state vector $\bm{s}$ is maintained by our agent, which describes the confidence about $o_{tgt}$.
At each time-step $t$, the agent picks up the promising action (select a question) according to the policy $\pi_\theta(a|\bm{s}_t)$, and transits from the state $\bm{s_t}$ to the next state $\bm{s_{t+1}}$ after receiving the answer (``Yes''/``No''/``Unknown'') from the user.
The historical trajectories $\langle s_t,a_t,r_{t+1},s_{t+1} \rangle$ are stored in a \emph{replay memory} which enables the agent to be trained on previously observed data by sampling from it.
Note that only when a guess is made about $o_{tgt}$ at the end of game can the agent receive a reward signal, which makes it unable to distinguish the importance of each selected question.
Therefore, we design a RewardNet to learn the more informative reward at each time-step and thus lead the agent to achieve the better performance.

In the rest of this section, we first describe how to formulate the Q20 game into a RL framework, and then introduce the RewardNet. Finally, we will demonstrate our training procedure in detail.

\subsection{Modeling of the Q20 Game}
\label{sec:modeling}
In the Q20 game, the goal of our agent is to figure out the object $o_{tgt}$ that the user thinks of at the beginning of game by asking 20 questions.
We formulate the process of question selection as a finite Markov Decision Process (MDP) which can be solved with RL.
A tuple $\langle$$\mathcal{S}$,\ $\mathcal{A}$,\ $\mathcal{P}$,\ $\mathcal{R}$,\ $\gamma$$\rangle$ is defined to represent the MDP, where $\mathcal{S}$ is the continuous state space, $\mathcal{A}=\{a_1,a_2,\cdots,a_m\}$ is the set of all available actions, $\mathcal{P}(S_{t+1}=s'|S_t=s,A_t=a)$ is the transition probability matrix, $\mathcal{R}(s,a)$ is the reward function and $\gamma\in[0,1]$ is the discount factor used to calculate the long-time return. 
In the RL framework, at each time-step $t$, the agent takes an action $a_t$ under the state $s_t$ according to the policy $\pi_\theta(a|s_t)$.
After interacting with the environment, the agent receives a reward scalar $r_{t+1}$ and transits to the next state $s_{t+1}$, then another time-step begins.
All these trajectories $\langle s_t,a_t,r_{t+1},s_{t+1} \rangle$ in a game constitute an episode which is an instance of the finite MDP.
The long-time return $G_t$ of the time-step $t$ is calculated as follows:
\begin{equation}
\label{eq:return}
G_t=\sum_{k=0}^{T}{\gamma^k r_{t+k+1}}
\end{equation}
In the following parts, we describe each component of RL corresponding to the Q20 game.

\cparagraph{Environment}
The major component of our environment is the user in the Q20 game who decides the target object $o_{tgt}$ and answers questions from the agent. 
Besides, the environment also needs to deliver the reward based on the outcome of the game and store historical data into the replay memory (see Fig.~\ref{fig:method_overview}).

\cparagraph{Action}
Since the agent interacts with the user by asking questions, the action $a_t\in{\mathcal{A}}$ taken by our agent refers to selecting the question $q_{a_t}$ at time-step $t$, and $\mathcal{A}$ is the set of the indices to all available questions in the Q20 game.

\cparagraph{State}
In our method, we use the state $\bm{s}_t$ to keep track of the current confidence of target object $o_{tgt}$.
Specifically $\bm{s}_t \in \mathbb{R^{|\mathcal{O}|}}$ and $\sum_{i=1}^{n}{s_{t,i}}=1$, where $\mathcal{O}=\{o_1,o_2,\cdots,o_n\}$ represents the set of all the objects that can be chosen by the user.
Therefore, the state $\bm{s}_{t}$ is a probability distribution over all the objects and $s_{t,i}$ is the confidence that the object $o_i$ is the target object $o_{tgt}$ at time-step $t$.

The initial state $\bm{s}_0$ can either be a uniform distribution or initialized by the prior knowledge.
We observe that users typically prefer to choose popular objects which are more concerned by the public.
For example, the founder of Tesla Inc. and the designer of SpaceX, ``Elon Musk'', is more likely to be chosen compared to a CEO of a new startup.
Motivated by this, we could use the yearly retrieval frequency $C(o_i)$ of object $o_i$ on a commercial search engine to calculate the initial state $\bm{s}_0$, where $s_{0,i}=C(o_i)\;/\;{\sum_{j=1}^{n}{C(o_j)}}$.

\cparagraph{Transition Dynamics}
In our method, the transition dynamics is deterministic.
Given the object set $\mathcal{O}$ and the question set $\mathcal{A}$, we collect the normalized probabilities of the answer over ``Yes'', ``No'' and ``Unknown'' for each object-question pair. And the rule of state transition is define as:
\begin{equation}
\label{eq:state_transfer}
\bm{s}_{t+1}=\bm{s}_{t} \odot \bm{\alpha}
\end{equation}
where $\bm{\alpha}$ depends on the answer $x_t$ to the question $q_{a_t}$ which is selected by the agent at the step $t$:
\begin{equation}
\alpha=\left\{
\begin{array}{rcl}
{[R(1,a_t),\ldots,R(|\mathcal{O}|,a_t)],} & {x_t=Yes} \\
{[W(1,a_t),\ldots,W(|\mathcal{O}|,a_t)],} & {x_t=No} \\
{[U(1,a_t),\ldots,U(|\mathcal{O}|,a_t)],} & {x_t=Unk}
\end{array} \right.
\end{equation}
where $\mathcal{O}$ is the object set and for each object-question pair $(o_{i},q_{j})$, $R(i,j)$ and $ W(i,j)$ are calculated as follows:
\begin{equation}
\label{eq:transition_matrix_1}
\begin{split}
R(i,j)&=\frac{C_{yes}(i,j)+\delta}{C_{yes}(i,j)+C_{no}(i,j)+C_{unk}(i,j)+\lambda} \\
W(i,j)&=\frac{C_{no}(i,j)+\delta}{C_{yes}(i,j)+C_{no}(i,j)+C_{unk}(i,j)+\lambda}
\end{split}
\end{equation}
$R(i,j)$ and $W(i,j)$ are probabilities of answering ``Yes'' and ``No'' to question $q_{j}$ with respect to the object $o_{i}$ respectively.
$C_{yes}(i,j)$, $C_{no}(i,j)$ and $C_{unk}(i,j)$ are frequencies of answering ``Yes'', ``No'' and ``Unknown'' to question $q_{j}$ with respect to the object $o_{i}$.
$\delta$ and $\lambda$ are smoothing parameters.
Then the probability of answering ``Unknown'' to question $q_{j}$ with respect to the object $o_{i}$ is:
\begin{equation}
\label{eq:transition_matrix_2}
U(i,j)=1-R(i,j)-W(i,j)
\end{equation}

In this way, the confidence $s_{t,i}$ that the object $o_i$ is the target object $o_{tgt}$ is updated following the user's answer $x_t$ to the selected question $q_{a_t}$ at the time-step $t$.

\cparagraph{Policy Network}
We directly parameterize the policy $\pi_\theta(a|\bm{s_t})$ with a neural network which maps the state $\bm{s_t}$ to a probability distribution over all available actions: $\pi_\theta(a|\bm{s_t})=\mathbb{P}[a|\bm{s_t};\theta]$. 
The parameters $\theta$ are updated to maximize the expected return which is received from the environment.
Instead of learning a greedy policy in value-based methods like DQN, the policy network is able to learn a stochastic policy which can increase the diversity of questions asked by our agent and potentially make the agent more robust to noisy answers in the real-world Q20 game.
The policy $\pi_\theta(a|\bm{s})$ is modeled by a Multi-Layer Perceptron (MLP) and the output layer is normalized by using a masked softmax function to avoid selecting the question that has been asked before.
Because asking the same question twice does not provide extra information about $o_{tgt}$ in a game.

\subsection{Problem of Direct Reward}
For most reinforcement learning applications, it is always a critical part to design reward functions, especially when the agent needs to precisely take actions in a complex task. A good reward function can improve the learning efficiency and help the agent achieve better performances.

In the Q20 game, however, the immediate reward $r_t$ of selecting question $q_{a_t}$ is unknown at the time-step $t$ $(t<T)$ because each selected question is just answered with a simple ``Yes'', ``No'' or ``Unknown'' and there is no extra information provided by user. 
Only when the game ends $(t=T)$ can the agent receive a reward signal of win or loss. So we intuitively consider the \textbf{\emph{direct reward}}: $r_T=30$ and $-30$ for the win and loss respectively while $r_t=0$ for all $t<T$.
Unfortunately, the direct reward is not discriminative because the agent receives the same immediate reward $r_t=0\,(t<T)$ for selecting both good and bad questions.
For example, if the $o_{tgt}$ is ``Donald Trump'', then selecting question (a) ``Is your role the American president?'' should receive more immediate reward $r_t$ than selecting question (b) ``Has your role been married?''.
The reason is that as for the $o_{tgt}$, question (a) is more relevant and can narrow down the searching space to a greater extent.

Therefore, it is necessary to design a better reward function to estimate a non-zero immediate reward $r_t$, and make the long-time return $G_t=\sum_{k=0}^{T}{\gamma^k r_{t+k+1}}$ more informative.

\subsection{Reward Function Approximation by Neural Network} 
\label{sec:rewardNet}
To solve the problem of the direct reward, we propose a reward function which employs a neural network to estimate a non-zero immediate reward $r_t$ at each time-step.
So that $G_t$ can be more informative, which thus leads to a better trained questioner agent.

The reward function takes the state-action pair $(s_t, a_t)$ as input and outputs the corresponding immediate reward $r_{t+1}$. 
In our method, we use a MLP with sigmoid output to learn the appropriate immediate reward during training, and this network is referred as \textbf{\emph{RewardNet}}.
In each episode, the long-term return $G_t$ is used as a surrogate indicator of $r_{t+1}$ to train our RewardNet with the following loss function:
\begin{equation}
\label{eq:reward_approx}
L_1(\sigma)=(R(s_t,a_t;\sigma)-\sigmoid(G_t))^2
\end{equation}
where $\sigma$ is the network parameters.
Here we apply the sigmoid function on $G_t$ so as to prevent $G_t$ from growing too large.
Besides, we also use the replay memory to store both old and recent experiences, and then train the network by sampling mini-batches from it.
The training process based on the experience replay technique can decorrelate the sample data and thus make the training of the RewardNet more efficient.

Furthermore, since the target object $o_{tgt}$ can be obtained at the end of each episode, we can use the extra information provided by $o_{tgt}$ to estimate a better immediate reward $r_t$.
To capture the relevance between the selected questions and $o_{tgt}$ in an episode, we further propose a \textbf{\emph{object-aware RewardNet}} which takes the $\langle s_t, a_t, o_{tgt} \rangle$ tuple as input and produces corresponding $r_{t+1}$ as output.
The detailed training algorithm is shown in Algo.~\ref{alg:rewardnet}.

\begin{algorithm}  
	\caption{Training Object-Aware RewardNet}  
	\label{alg:rewardnet}   
	Initialize replay memory $\mathcal{D}_1$ to capacity $N_1$ \\
	Initialize RewardNet with random weights $\sigma$ \\
	\For{$episode\ i \gets 1$ \textbf{to} $Z$}
	{
		User chooses object $o_i$ from $\mathcal{O}$ \\
		Initialize temporary set $S_1$ and $S_2$ \\
		Play with policy $\pi_{\theta}(a_t|s_t)$, and store $(s_t,a_t)$ in $S_1$, where $t\in[0, T]$\\
		$r_T \gets 30$ or $-30$ for a win or loss\\
		\For{$(s_t,a_t)$ \textbf{in} $S_1$}
		{
			Get $r_{t+1}$ from \textbf{RewardNet} \\
			Store $(s_t,a_t, r_{t+1})$ tuple in $S_2$ \\
		}
		\For{$(s_t,a_t,r_{t+1})$ \textbf{in} $S_2$}
		{
			$G_t \gets \sum_{k=0}^{T}{\gamma^k r_{t+k+1}}$ \\
			$r'_{t+1} \gets sigmoid(G_t)$\\
			Store $(s_t,a_t,o_i,r'_{t+1})$ in $\mathcal{D}_1$ \\
			\If{ $len(\mathcal{D}_1) > K_1$ }
			{
				Sample mini-batch from $\mathcal{D}_1$ \\
				Update $\sigma$ with loss $L_1(\sigma)$ in Eq.~\ref{eq:reward_approx} \\
			}
		}
	}
\end{algorithm}

\subsection{Training the Policy-Based Agent}
We train the policy network using REINFORCE~\cite{williams1992simple} algorithm and the corresponding loss function is defined as follows:
\begin{equation}
\label{eq:policyNetLoss}
L_2(\theta)=-\mathbb{E}_{\pi_\theta}[\log{\pi_\theta(a_t|s_t)(G_t-b_t)}]
\end{equation}
where the baseline $b_t$ is a estimated value of the expected future reward at the state $s_t$, which is produced by a value network $\mathcal{V}_{\eta}(s_{t})$. 
Similarly, the value network $\mathcal{V}_{\eta}(s_{t})$ is modeled as a MLP which takes the state $s_t$ as input and outputs a real value as the expected return.
By introducing the baseline $b_t$ for the policy gradient, we can reduce the variance of gradients and thus make the training process of policy network more stable.
The network parameters $\eta$ are updated by minimizing the loss function below:
\begin{equation}
\label{eq:valueNetLoss}
L_3(\eta)=(\mathcal{V}_{\eta}(s_{t})-G_t)^2
\end{equation}

Note that, in our method, both the RewardNet and the value network $\mathcal{V}_{\eta}(s_{t})$ approximate the reward during training.
But the difference lies in that the RewardNet is designed to estimate a appropriate non-zero reward $r_t$ and further derive the more informative return $G_t$ while $\mathcal{V}_{\eta}(s_{t})$ aims to learn a baseline $b_t$ to reduce the variance of policy gradients. 
We combine both of two networks to improve the gradients for our policy network and thus lead to a better agent.
The training procedure is described in Algo.~\ref{alg:drl}.

\begin{algorithm}  
	\caption{Training the Agent}  
	\label{alg:drl}   
	Initialize replay memory $\mathcal{D}_2$ to capacity $N_2$ \\
	Initialize policy net $\pi$ with random weights $\theta$ \\
	Initialize value net $\mathcal{V}$ with random weights $\eta$ \\
	Initialize RewardNet with random weights $\sigma$ \\
	\For{$episode\ i \gets 1$ \textbf{to} $Z$}
	{
		Rollout, collect rewards, and save the history in $S_2$ (4-10 in Algo.~\ref{alg:rewardnet})\\
		\For{$(s_t,a_t,r_{t+1})$ \textbf{in} $S_2$}
		{
			$G_t \gets \sum_{k=0}^{T}{\gamma^k r_{t+k+1}}$ \\
			Update \textbf{RewardNet} (13-17 in Algo.~\ref{alg:rewardnet}) \\
			Store $(s_t,a_t,G_t)$ in $\mathcal{D}_2$ \\
			\If{ $len(\mathcal{D}_2) > K_2$ }
			{
				Sample mini-batch from $\mathcal{D}_2$ \\
				Update $\eta$ with loss $L_3$ in Eq.~\ref{eq:valueNetLoss}\\
				Update $\theta$ with loss $L_2$ in Eq.~\ref{eq:policyNetLoss}\\
			}
		}
	}
\end{algorithm}

\section{Experimental Setup}
We use a user simulator to train our questioner agent and test the agent with the simulated answerer and real users.
Specifically, our experiments answer three questions:
(1) Is our method more robust in real-world Q20 games, compared to the methods based on relevance table? (Section.~\ref{sec:human_eval})
And how does it perform in the simulation environment? (Section.~\ref{sec:simulation_eval})
(2) Does our RewardNet help in the training process? (Section.~\ref{sec:exp_reward})
(3) How the winning rate grows with the number of questions, and whether it is possible to stop earlier? (Section.~\ref{sec:q_num})

\subsection{User Simulator}
\label{sec:sumulator}
Training the RL agent is challenging because the agent needs to continuously interact with the environment.
To speed up the training process of the proposed RL model, we construct a user simulator which has enough prior knowledge to choose objects and answer questions selected by the agent.

We collect 1,000 famous people and 500 questions for them.
Besides, for every person-question pair in our dataset, a prior frequency distribution over ``Yes'', ``No'' and ``Unknown'' is also collected from thousands of real users.
For example, as for ``Donald Trump'', question (a) ``Is your role the American president?'' is answered with ``Yes'' for 9,500 times, ``No'' for 50 times and ``Unknown'' for 450 times.
We use Eq.\ref{eq:transition_matrix_1} and \ref{eq:transition_matrix_2} to construct three matrices $R,W,U \in \mathbb{R^{|\mathcal{O}|*|\mathcal{A}|}}$ ($|\mathcal{O}|=1000,|\mathcal{A}|=500$) which are used for state transition in the Section.~\ref{sec:modeling}. Then given the object $o_i$ and question $q_j$, the user simulator answers  ``Yes'', ``No'' and ``Unknown'' when $R(i,j)$, $W(i,j)$, and $U(i,j)$ has the max value among them respectively.

Constructed by the prior knowledge, the simulator can give noise-free answer in most cases. 
Because the prior frequency distribution for each person-question pair is collected from thousands of users with the assumption that most of them do not lie when answering questions in the Q20 game.

In an episode, the simulator randomly samples a person following the object distribution $\bm{s}_0$, which is generated from the object popularity (see the state part of Section.~\ref{sec:modeling}), as the target object. 
Then the agent gives a guess when the number of selected questions reaches 20. 
After that, the simulator check the agent's answer and return a reward signal of win or loss.
There is only one chance for the agent to guess in an episode. The win and loss reward are 30 and -30 respectively.

\subsection{Implementation Details}
While the architectures of the policy network, RewardNet and value network can vary in different scenarios, in this paper, we simply use the MLP with one hidden layer of size 1,000 for all of them, but with different parameters.
These networks take in the state vector directly, which is a probability distribution over all objects.
The RewardNet further takes in the one-hot vector of action $a_t$.
Based on the input of RewardNet, the object-aware RewardNet takes one more target object $o_{tgt}$ as the feature which is also a one-hot vector.

We use the ADAM optimizer~\cite{kingma2014adam} with the learning rate 1e-3 for policy network and 1e-2 for both RewardNet and value network.
The discounted factor $\gamma$ for calculating the long-term return is 0.99.
The model was trained up to 2,000,000 steps (2,00,000 games) and the policy network was evaluated every 5,000 steps.
Each evaluation records the agent's performance with a greedy policy for 2,000 independent episodes.
The 2,000 target objects for these 2,000 episodes are randomly selected following the distribution $\bm{s}_0$, which is generated from the object popularity and kept the same for all the training settings.

\subsection{Competitor}
\label{sec:baseline}
We compare our RL method with the entropy-based model proposed by Wu et al.~\shortcite{wu2018q20}, which utilizes the real-world answers to each object-question pair to calculate an object-question relevance matrix with the entropy-based method.
The relevance matrix is then used for question ranking and object ranking via carefully designed formulas and engineering tricks.
Since this method is shown to be effective in their production environment, we consider it to be a strong baseline to our proposed RL model.

\section{Experimental Results}
\label{sec:exp_res}
\subsection{Simulated Evaluation}
\label{sec:simulation_eval}

\begin{figure}[!t]
	\centering
	\includegraphics[width=0.48\textwidth]{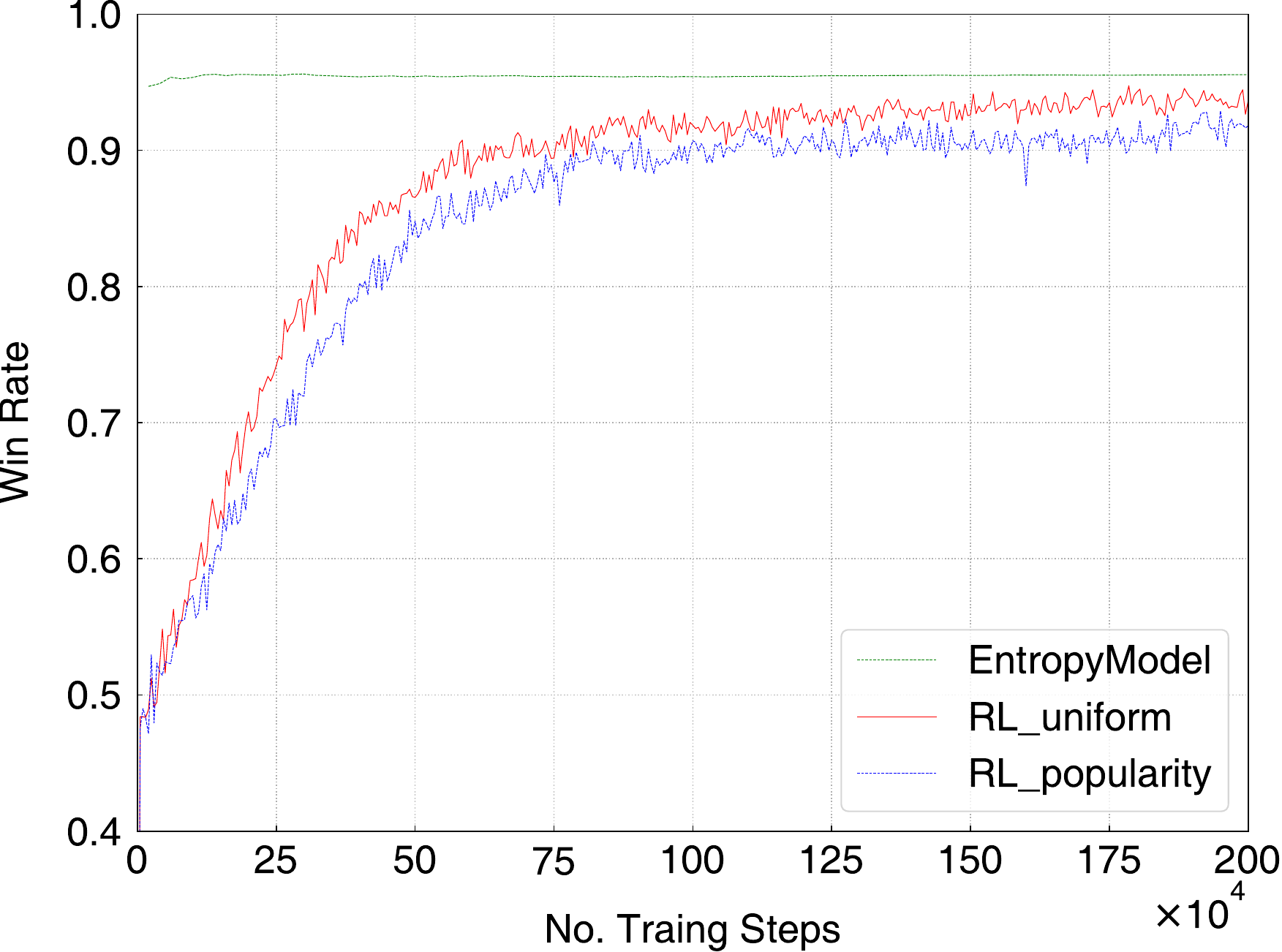}
	\caption{Win Rate Curves in Simulation Environment.}
	\label{fig:simulation_winrate}
\end{figure}

We first evaluate our agent and the entropy-based baseline (referred to as \entropy, see Section.~\ref{sec:baseline}) by using the simulated user (Section.~\ref{sec:sumulator}).
To investigate which initialization strategy of the state $\bm{s}_0$ is better (see the state part of Section.~\ref{sec:modeling}), we further evaluate two variants of our model: the agent with uniform distribution $\bm{s}_0$ (\RLunif) and the agent with the distribution $\bm{s}_0$ initialized by the prior knowledge on the object popularity (\RLpop).

Fig.~\ref{fig:simulation_winrate} shows the curves on the win rate of these methods evaluated on 2,000 independent episodes with respect to the number of training steps.
Note that, the \entropy only needs to update its statistics during training and has already accumulated a significant number of data since it has been run for over a year in their production environment.
Therefore, only a small fraction of its statistics can be changed, which leads to a small rise at the beginning of training, and its win rate remains at around 95\% afterwards.

On the other hand, both our RL models continuously improve the win rate with the growing number of interactions with the user simulator, and they achieve 50\% win rate after around 20,000 steps.
As we can see, although the $\bm{s}_0$ initialized with the prior knowledge of object popularity keeps consistent with the object selection strategy of the simulator, the agent with uniform distribution $\bm{s}_0$ (\RLunif) still performs clearly better than the agent with $\bm{s}_0$ based on the prior knowledge (\RLpop).
The reason is that the former can explore the Q20 game environment more fully.
The prior knowledge based $\bm{s}_0$ helps the agent narrow down the candidate space more quickly when the target object is a popular object.
However, it also becomes misleading when the target object is not popular and makes the agent even harder to correct the confidence of the target object.
On the contrary, the uniform distribution $\bm{s}_0$ makes the agent keep track of the target object only based on the user's answers.
And the superior performance of the \RLunif indicates that our question selection policy is highly effective, which means it is not necessary to use the \RLpop to increase the win rate of hot objects in the game.

As shown in Fig.~\ref{fig:simulation_winrate}, \RLunif achieves win rate 94\% which is very close to \entropy.
Compared to our RL method, \entropy needs more user data to calculate their entropy-based relevance matrix and involves many engineering tricks.
The fact that \RLunif is competitive to \entropy in the noise-free simulation environment indicates that our RL method is very cost-effective: it makes use of user data more efficiently and is easier to implement.

\subsection{Human Evaluation}
\label{sec:human_eval}
\begin{table}
	\begin{center}
		\begin{tabular}{lccc}
			\toprule
			& Win Rate \\
			\midrule
			\entropy & 71.3\% \\
			\RLunif & 75.9\% \\
			\bottomrule
		\end{tabular}
	\end{center}
	\caption{Win Rate on Human Evaluation.}
	\label{tab:human-evaluation}
\end{table}
To further investigate the performance of our RL method in the real-world Q20 game where noisy answers are common, we also conduct an human evaluation experiment.
Specifically, we let real users to play the game with \entropy and \RLunif for 1,000 times respectively.
In the real-world Q20 game, users sometimes make mistakes when they answer the questions during the game.
For example, as for the target object ``Donald Trump'', question (a) ``Is your role the American president?'' is sometimes answered with ``No'' or ``Unknown'' by real users.
On the contrary, the simulator hardly makes such mistakes since we have provided it with enough prior knowledge.
As shown in Table.~\ref{tab:human-evaluation}, \RLunif outperforms \entropy by about 4.5\% on win rate in the real-world Q20 games.
It shows that our RL method is more robust to noisy answers than \entropy.
Specifically, the robustness of our RL method to the noise is shown in the following two aspects.
First, compared to the rule-based statistics update in \entropy, our RL model can be trained by modern neural network optimizers in a principled way, which results in the better generalization ability of our model.
Secondly, different from the \entropy selecting the top-ranked question at each time-step, \RLunif samples a question following its question probability distribution $\pi_\theta(a|\bm{s})$, which enables our agent to jump out of the local optimum caused by incorrect answers from users.
And since more randomness is introduced by sampling from the question probability distribution during training, it also improves the tolerance of our model towards the unexpected question sequences. 

Besides, we also find some interesting cases during human evaluation.
Sometimes, the RL agent selects a few strange questions which seems to be not that much relevant to the chosen object, but it can still find the correct answer at the end of game.
This situation is caused by the fact that our method samples questions based on the output of policy net, rather than greedy selection during training. We find that this phenomenon increases the user experience since it makes the agent more unpredictable to the users.

\subsection{The Effectiveness of RewardNet}
\label{sec:exp_reward}
\begin{figure}[!t]
	\centering
	\includegraphics[width=0.48\textwidth]{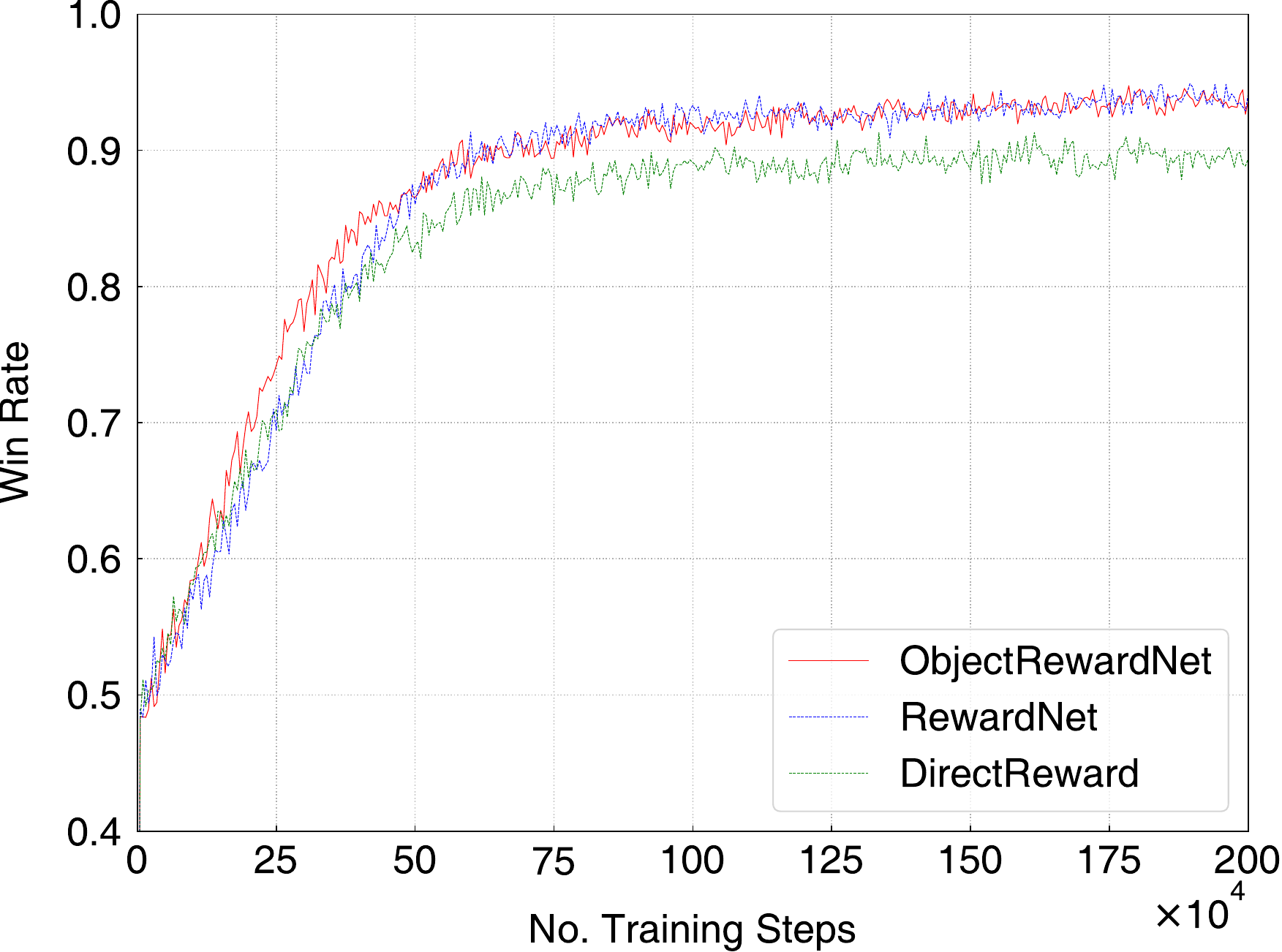}
	\caption{Effectiveness of RewardNet.}
	\label{sec:perf_reward} 
\end{figure}

To investigate the effectiveness of our RewardNet (Section.~\ref{sec:rewardNet}), we further evaluate three variants of our model in the simulation environment: the model trained with with direct reward, RewardNet, and object-aware RewardNet, which are referred to as \directR, \RN, and \ORN respectively. They are all trained with the uniform distribution $\bm{s}_0$.

As shown in Fig.~\ref{sec:perf_reward}, \directR converges in the early steps and has a relatively poor performance with the win rate 89\%.
Both \RN and \ORN achieve the better performance with a win rate of 94\% after convergence.
This clear improvement shows that the more informative long-term return, calculated with the immediate reward delivered by our RewardNet method, significantly helps the training of the agent.

Furthermore, as shown in Fig.~\ref{sec:perf_reward}, we can also see that \ORN learns faster than \RN in the early steps.
This indicates that \ORN can estimate the immediate reward more quickly with the extra information provided by the target object, which leads to the faster convergence of the agent.

\subsection{Win Rate Regarding Question Numbers}
\label{sec:q_num}

\begin{figure}[!t]
	\centering
	\includegraphics[width=0.48\textwidth]{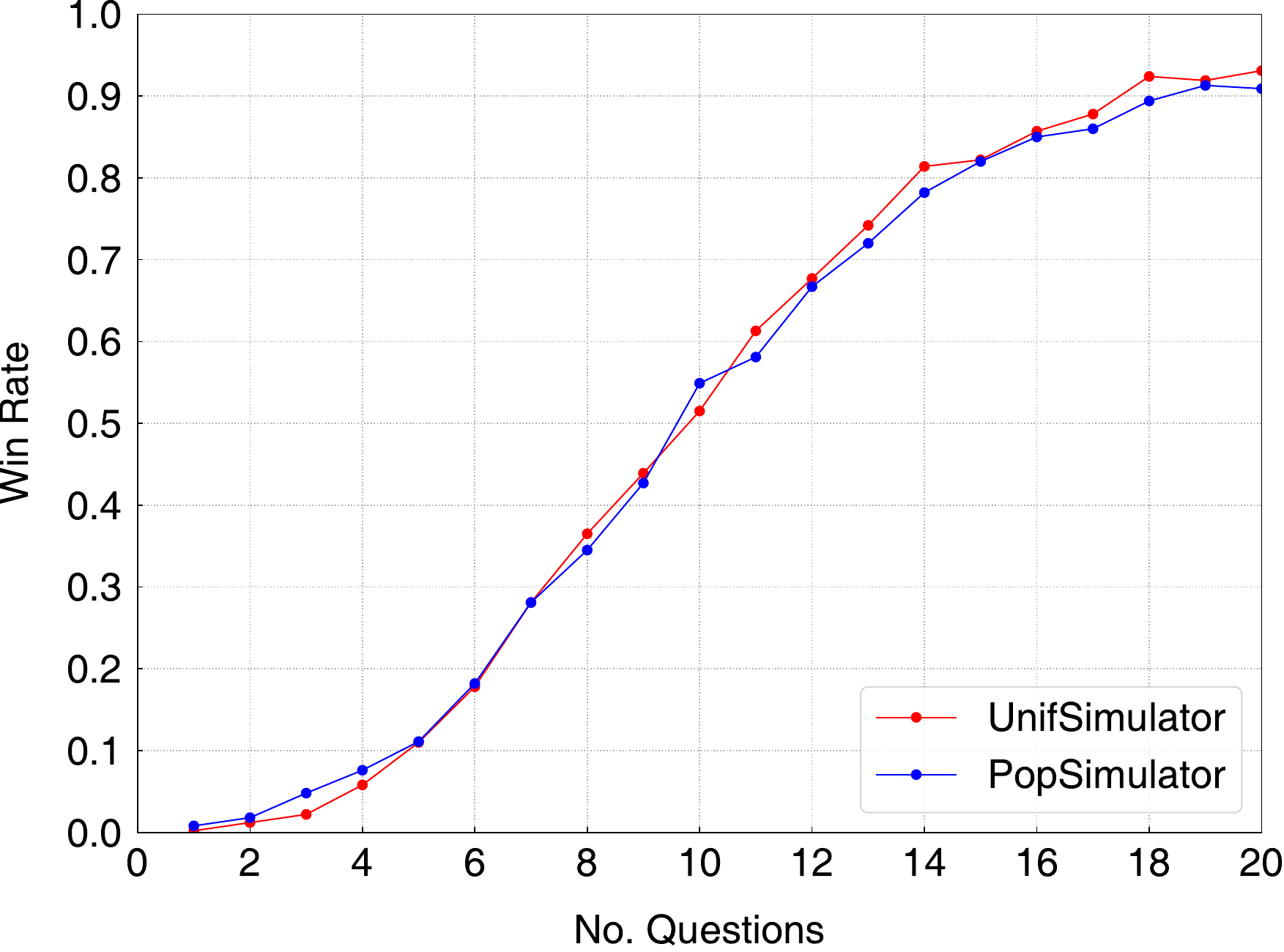}
	\caption{Win Rate Regarding Numbers of Questions.}
	\label{fig:perf_numQ}
\end{figure}

In this section, we investigate how the win rate grows with the number of asked questions and whether a early-stop strategy can be adopted in the game.
We use the user simulator to play the game with the \RLunif agent and two settings are taken into account:
the simulator samples the target object following the uniform object distribution (\unifSimu), and samples following the prior object distribution based on the object popularity (\popSimu). We perform 1,000 simulations for each number of questions, and the win rate curve is shown in Fig.~\ref{fig:perf_numQ}.

As we can see that \unifSimu achieves the win rate of 80\% with only 14 questions in both settings. 
And the flat curves in the region after 18 questions indicate that the game can be early stopped with the almost same win rate at step 18.
Since a lower win rate is acceptable sometimes, other early-stop strategies can also be derived for the better user experience with the trade-off between the win rate and game steps.

Besides, the fact that \RLunif performs similarly under both settings actually shows that our RL method is robust to different objects.
It also performs well on infrequent objects where we may have the limited user data for constructing a well-tuned state transition dynamics.

\subsection{Case Study}
When our agent is playing the game with real users, we select two cases from records.
In the first case, the person that the user chooses is Cristiano Ronaldo, the famous football player.
As we can see in Tab.~\ref{tab:user-case1}, our agent can still figure out the target person while No.17 and No.19 questions are answered wrong by the user, which indicates our agent is robust to noisy answers.
In the second case, the chosen person is Napoleon Bonaparte who was the French Emperor.
Although there are some other candidates satisfied the constraints, the target person can be figured out because of the people popularity, which is shown in Tab.~\ref{tab:user-case2}.
\section{Related Work}

\cparagraph{Q20}
The Q20 game is popularized as an electronic game by the program of Robin Burgener in 1988~\cite{Burgener2006Artificial}, which uses a object-question relevance table to rank questions and target objects.
Wu et al.~\cite{wu2018q20} improves the relevance table with entropy-based metrics, and uses complicated engineering tricks to make it perform quite well in their production environment.
These table-based methods use rules to update parameters, which makes them easily affected by noisy answers.
Besides, Zhao and Maxine~\shortcite{zhao2016towards} also explores Q20 in their dialogue state tracking research.
However, they only use a small toy Q20 setting where the designed questions are about 6 person attributes in the Knowledge Base (KB).
Since their method relies on the KB for narrowing down the scope of target object, it is not applicable to real-world Q20 games where a well-defined object KB is often unavailable.
Compared to previous approaches, our RL method is robust to the answer noise and does not rely on the KB.

\cparagraph{Deep Reinforcement Learning}
DRL has witnessed great success in playing complex games like Atari games~\cite{Mnih2015Human} , Go~\cite{silver2016mastering}, and etc.
In the natural language processing (NLP), DRL is also used to play text-based games~\cite{narasimhan2015language},
and used to handle fundamental NLP tasks like machine translation~\cite{he2016dual} and machine comprehension~\cite{hu2017reinforced} as well.
Our Q20 game lies in the intersection of the field of game and NLP.
In this work, we propose a policy-based RL model that acts as the questioner in the Q20 game, and it exhibits the superior performance in our human evaluation.

\cparagraph{Natural Language Games}
In the literature, there are some works focusing on solving and generating English riddles~\cite{de1992riddles,binsted1996machine} and Chinese character riddles~\cite{tan2016solving}.
Compared to riddles, the Q20 game is a sequential decision process which requires careful modeling of this property.

\section{Conclusions}
In this paper, we propose a policy-based RL method to solve the  question selection problem in the Q20 Game. 
Instead of using the direct reward, we further propose an object-aware RewardNet to estimate the appropriate non-zero reward and thus make the long-time return more informative.
Compared to previous approaches, our RL method is more robust to the answer noise which is common in the real-world Q20 game.
Besides, our RL agent can also ask various questions and does not require the existing KB and complicated engineering tricks.
The experiments on a noisy-free simulation environment show that our RL method is competitive to an entropy-based engineering system, and clearly outperforms it on the human evaluation where noisy answers are common.

As for the future work, we plan to explore  methods to use machine reading to automatically construct the state transition dynamics from corpora like Wikipedia.
In this way, we can further build an end-to-end framework for the large-scale Q20 games in the real world. 
\section*{Acknowledgement}
We gratefully thank the anonymous reviewers for their insightful comments and suggestions on the earlier version of this paper.
The first author also thanks the Microsoft for providing resources for the research.

\bibliography{emnlp2018}
\bibliographystyle{acl_natbib_nourl}

\clearpage

\appendix


\begin{table*}[t]
    \centering
    \resizebox{0.75\linewidth}{!} {
        \begin{tabular}{lll}
            \toprule
            No. &  Question & User's Answer  \\ \midrule
            1  &  Was the person born in Asia?   & No  \\
            2  &  Is the person very famous?   & Yes  \\
            3  &  Is the person a actor or actress?   & No  \\
            4  &  Is the person still alive?  & Yes  \\
            5  &  Was the person born in the 1990s?  & No \\
            6  &  Is the person the founder of a famous company?  & No \\
            7  &  Did the person finish the college in USA?  & No \\
            8  &  Is the person a famous singer?  & No \\
            9  &  Is the person male?  & Yes \\
            10 &  Is the person related to sports?  & Yes \\
            11 &  Is the person a football player?  & Yes \\
            12 &  Is the person a midfielder?  & No \\
            13 &  Is the person played for a European football club?  & Yes \\
            14 &  Is the person playing in the Spanish Premier League?  & Yes \\
            15 &  Is the person famous for the handsome or beautiful looks?  & Yes \\
            16 &  Does the person have big muscles?  & Yes \\
            17 &  Does the person have brown hair?  & No \\
            18 &  Will you be happy when you see the person?  & Yes \\
            19 &  Does the person have brothers or sisters in the family?  & No \\
            20 &  Is the person engaged in many charity activities?  & Unknown \\ 
            \bottomrule
        \end{tabular}
    }
	\caption{The person that the user chooses is Cristiano Ronaldo, the famous football player. As we can see in table, our agent can still figure out the target person while No.17 and No.19 are answered wrong by the user, which indicates our agent is robust to noisy answers.}
	\label{tab:user-case1}
\vspace{10mm}
	\resizebox{0.75\linewidth}{!} {
    	\begin{tabular}{lll}
            \toprule
            No. &  Question & User's Answer  \\ \midrule
            1  &  Is the person female?  & No \\
            2  &  Is the person still alive?  & No \\
            3  &  Does the person have children?   & Yes \\
            4  &  Does the person have brothers or sisters in the family?  & Yes \\
            4  &  Is the person very smart?  & Yes \\
            5  &  Was the person born in America?  & No \\
            6  &  Is the person the white man?  & Yes \\
            7  &  Is the person's family very rich?  & No \\
            8  &  Is the person a controversial figure in history?  & Yes \\
            9  &  Is the person related to politics?  & Yes \\
            10 &  Does the person have good looks?  & Unknown \\
            11 &  Does the person have short hair?  & Yes \\
            12 &  Is the person very famous?  & Yes \\
            13 &  Has the person once been very powerful?  & Yes \\
            14 &  Is the character of the person very aggressive?  & No \\
            15 &  Has the person been the president of a country?  & Yes \\
            16 &  Is the person a military?  & Yes \\
            17 &  Has the person once killed men?  & No \\
            18 &  Was the person born in Britain?  & No \\
            19 &  Was the person one of famous leaders in the World War II?  & No \\
            20 &  Has the person once been the emperor?  & Yes \\
            \bottomrule
        \end{tabular}
    }
    \caption{In this case, the person that the user chooses is Napoleon Bonaparte, the French Emperor. Although there are some other candidates satisfied the constraints, our agent can figure out the target person because of the people popularity.}
	\label{tab:user-case2}
	
\end{table*}

\end{document}